\documentclass[aps,prl,reprint,superscriptaddress, longbibliography]{revtex4-1}
\usepackage{amsmath,amssymb,graphicx}
\usepackage[colorlinks=true]{hyperref}

\usepackage{hyphenat}
\usepackage{graphicx}

\begin{document}

\title{Reciprocal space engineering with hyperuniform gold metasurfaces}

\author{Marta \surname{Castro-Lopez}}
\author{Michele \surname{Gaio}}
\affiliation{Department of Physics, King's College London, Strand, London WCR
2LS, United Kingdom.}

\author{Steven \surname{Sellers}}
\author{George \surname{Gkantzounis}}
\author{Marian \surname{Florescu}}
\email{m.florescu@surrey.ac.uk}
\affiliation{Department of Physics, University of Surrey, Guildford, Surrey GU27XH,
United Kingdom.}

\author{Riccardo \surname{Sapienza}}
\email{riccardo.sapienza@kcl.ac.uk}
\affiliation{Department of Physics, King's College London, Strand, London WCR
2LS, United Kingdom.}

\begin{abstract}
Hyperuniform geometries feature correlated disordered topologies which follow
from a tailored k-space design. Here we study gold plasmonic hyperuniform
metasurfaces and we report evidence of the effectiveness of k-space engineering on both light scattering
and light emission experiments. The metasurfaces possess interesting
directional emission properties which are revealed by momentum spectroscopy as diffraction
and fluorescence emission rings at size-specific k-vectors. The opening of these rotational-symmetric
patterns scales with the hyperuniform correlation length parameter as
 predicted via the spectral function method.
\end{abstract}

\maketitle

Coherent control of optical waves by scattering from 2D nanostructured
surfaces is revolutionising the way we shape the wavefront of an incoming
light beam, opening new avenues for miniaturised optical components for integrated optical circuits~\cite{Gholipour2013},
flat display technology~\cite{Hosseini2014}, and energy harvesting~\cite{Azad2016,Almoneef2015}.
Metallic surfaces are in particular attractive due to the strong light-matter interaction associated
with surface plasmons, enabling diffraction control through plasmonic
crystals~\cite{Barnes1995,Bouillard2014} and metal nano-particle
arrays~\cite{Giannini2010,DalNegro2012}, broadband operation and
increase of the plasmon mode density~\cite{Lubin2013}, enhanced omni-directional
light extraction and coupling~\cite{Lawrence2012}, broadband absorption~\cite{Afshinmanesh2014},
fluorescence enhancement~\cite{Gaio2015a}
and lasing~\cite{Schokker2015,Zhang2016}, and more recently the realisation
of ultra thin lenses~\cite{Capasso2011} and metasurface holograms~\cite{Ni2013}.

Whereas periodic geometries suffer from limited rotational symmetries,
aperiodic and disordered topologies, with their richer symmetries and
patterns, can lead to superior optical functionalities~\cite{DalNegro2012},
as in omnidirectional absorption for solar applications~\cite{Martins2013,Burresi2013},
scattering-induced light localisation~\cite{Lagendijk2009} and light
extraction from LED/OLED~\cite{Koo2010}. Moreover, disordered metasurfaces are expected
to be more resilient against fabrication imperfection and therefore
more apt for technological implementation. Given the vast possible
designs of non-periodic topologies, ranging from random to correlated-disordered,
their full potential is still to be fully explored.

There exists a general class of disordered systems, called hyper\-uniform
disordered (HuD) photonic structures, which are of particular interest because they exhibit wide and isotropic
photonic band gaps~\cite{FroufePRL}, rotational symmetry and broadband k-space
control, and can be systematically generated through a specific design
rule via universal tessellation protocol~\cite{Florescu2009, Florescu2009b}. Pioneering
experiments on photonic HuD systems have explored IR light diffraction in
3D dielectric structures~\cite{Muller2014}, microwave band-gaps formation~\cite{Man2013},
polarization filtering~\cite{Zhou2016} and random quantum cascade lasers~\cite{DeglInnocenti2016}.
Theoretical proposals have been put forward for surface enhanced
Raman scattering~\cite{DeRosa2015}, transparency design~\cite{Leseur2016},
high-Q optical cavities and low-loss waveguides~\cite{PhysRevB.87.165116, Amoah2015, Band2016},
and microwave photonic circuits~\cite{Man2013}. HuD structures
fabrication is improving quickly, reaching already the IR range~\cite{Muller2014} but not
yet the visible.

Here, we report visible light scattering and light emission experiments
from HuD plasmonic gold metasurfaces. We find that scattering from
the metasurfaces is principally directed into an annular
angular pattern indicating reciprocal space engineering. Moreover,
we investigate directional emission from near-field coupled emitters
 which, as confirmed by theoretical modelling, is shaped into a ring from the
effective band folding into the light cone by scattering processes.

\begin{figure}[!htb]
\centering\includegraphics[width=7cm]{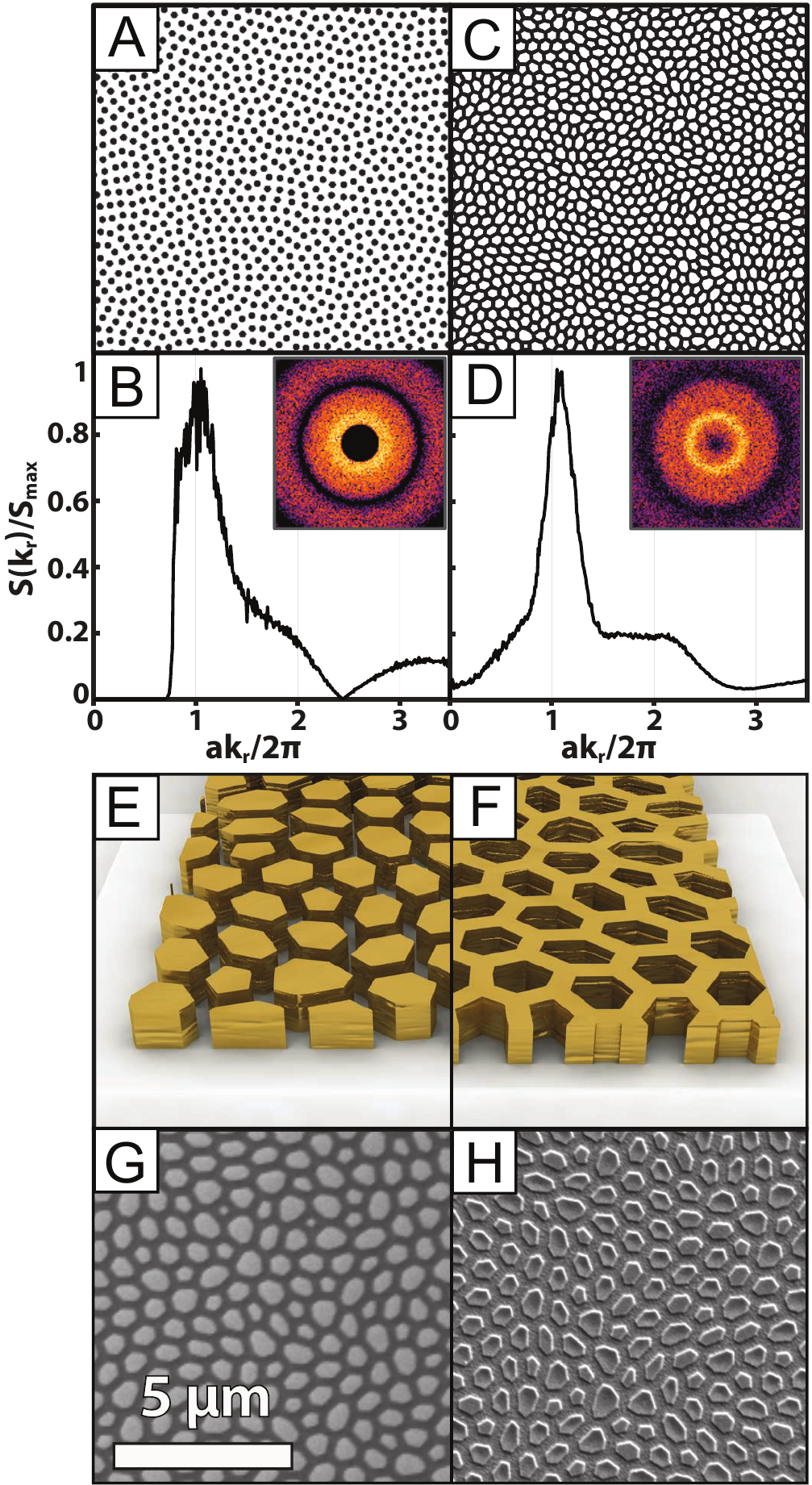}
\caption{The initial HuD
point pattern (panel A) with $\chi=0.49$ presents a structure factor (panel B)
with a typical zero around $\mathbf{k}=0$ and a broad isotropic diffraction
maximum (inset). The HuD connected network (panel C) broadly preserves the $\mathbf{k}$
space characteristics (panel D) of the point pattern. Sketches and SEM images of
the resulting pillar-type (panel E and G)
and network-type (panel F and H) metasurfaces.}
\label{FIG:KingsDesigns}
\end{figure}

The metasurface design is derived from a stealth hyperuniform point
pattern with  $\chi=0.49$ comprising $4000$ points~\cite{Florescu2009} and
generated under a periodic boundary condition, for a given average inter-scatterer distance
$a$. A section of the point pattern, decorated by discs
of radius $0.3a$, is shown in Fig.~\ref{FIG:KingsDesigns}A. The
structure factor of the point pattern is shown in Fig.~\ref{FIG:KingsDesigns}B:
the point distribution exhibits significant local structural correlations:
the typical exclusion region around $\mathbf{k}=0$ which characterizes
stealthy hyperuniform patterns and a broad isotropic diffraction maximum
peaked around $ak/2\pi=1.03$.

Next, a Delaunay tessellation protocol~\cite{Florescu2009}
is performed,  obtaining a strictly trivalent continuous network
topology (Fig.~\ref{FIG:KingsDesigns}C) with walls of thickness
$0.35a$. The structure factor of this network is presented
in Fig.~\ref{FIG:KingsDesigns}D. The stealthiness of the architecture
has been significantly reduced: the diffraction spectrum (Fig.~\ref{FIG:KingsDesigns}D,
inset) exhibits low intensity diffuse scattering around $\mathbf{k}=0$,
in contrast to the sharp exclusion zone of the simple point pattern
(Fig.~\ref{FIG:KingsDesigns}B, inset). Nonetheless, the general
form of the point pattern structure factor dominated by a single broad and isotropic
resonance around $ak/2\pi=1.09$ is reproduced by the network.

Gold metasurfaces were fabricated by electron beam lithography on
a glass substrate for various size scaling
parameters $a$. The samples are of two kind:  \emph{pillar-type} samples comprising isolated
pentagonal, hexagonal and heptagonal gold pillars (sketched in Fig.~\ref{FIG:KingsDesigns}E and SEM in Fig.~\ref{FIG:KingsDesigns}G),
 and \emph{network-type} designs (identical but inverted) consisting
of a connected network of gold (sketch in Fig.~\ref{FIG:KingsDesigns}F and SEM in Fig.~\ref{FIG:KingsDesigns}H).
 The samples are labelled
as $L_pN$ for pillars and $L_nN$ for networks, with  $N=a\times\sqrt{4000}$: for example $L_{p}50$ for $a =790$\,nm
has a side dimension of $50\,\mu$m.
 Network designs larger or smaller than $40\,\mu$m have been cropped or periodically
repeated, respectively, to cover a $40\times40\,\mu$m$^{2}$ area.

\begin{figure*}[htb]
\centering
\includegraphics[width=13cm]{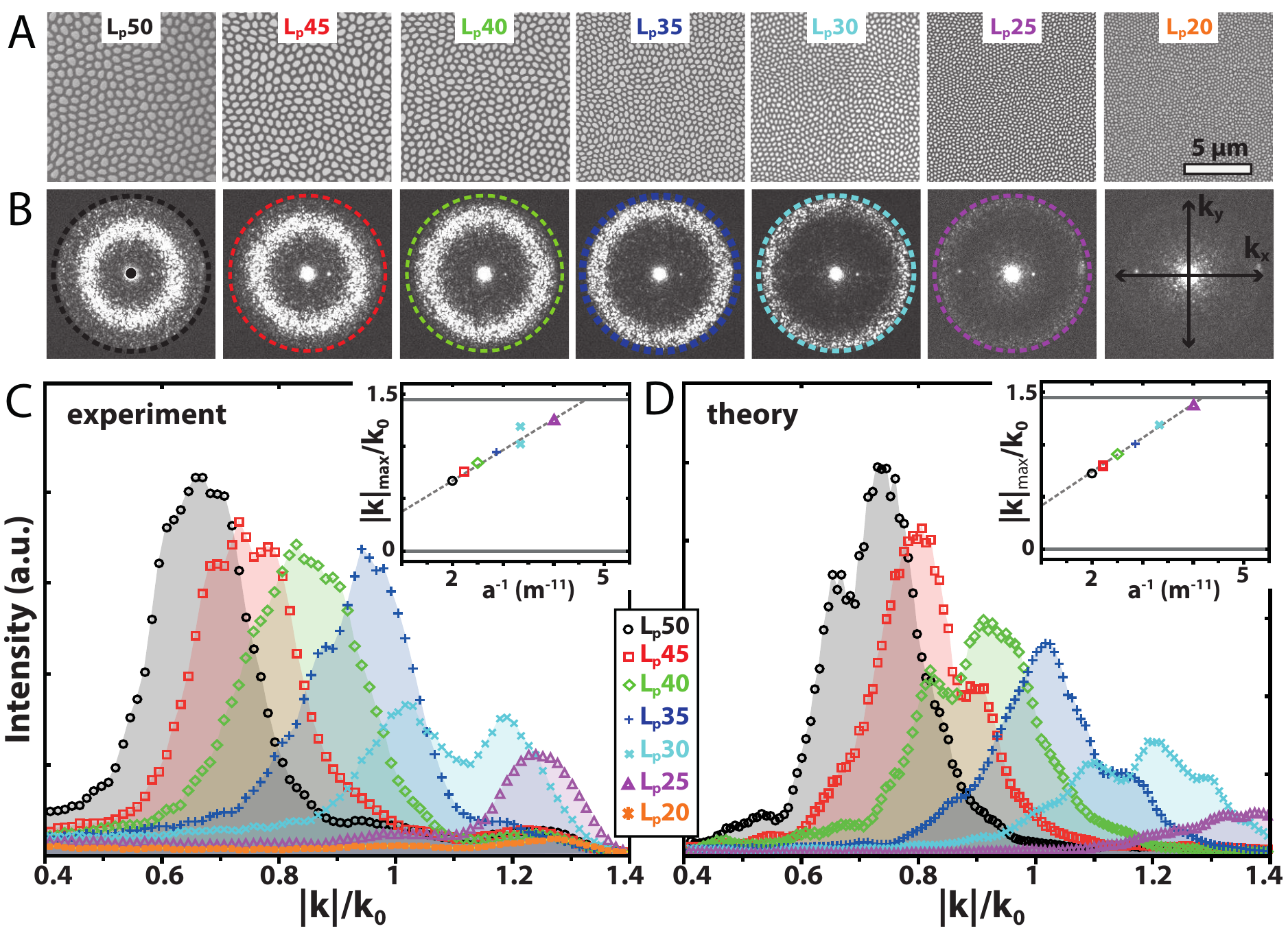}\caption{
SEM images of pillar type designs (panel A) together
with their farfield diffraction patterns (panel B) when illuminated
with a $532$~nm laser. The numerical aperture limits are marked as coloured
rings. Panel C displays the azimuthally integrated farfields distributions  as a function
of in-plane momentum (normalized to the incident wavevector $k_{0}$),
showing broad scattering resonances.
Panel D shows the same calculated azimuthally integrated farfields which
 agrees well with experiment. The insets (panel C \& D) plots
the scattering peak position as a function of the reciprocal scaling parameter ($a^{-1}$)
and the expected linear dependence.}
 \label{FIG:KingsDiffraction}
\end{figure*}

Fig.~\ref{FIG:KingsDiffraction} presents light scattering experiments on the pillar-type metasurfaces
whose SEM images are shown in Fig.~\ref{FIG:KingsDiffraction}A. Similar
experiments performed on the network-type samples  led to comparable
results and are not shown here.
The measured farfield intensity distribution of each pillar metasurface
is shown in Fig.~\ref{FIG:KingsDiffraction}B. The samples were illuminated
through the glass substrate with a collimated laser ($\lambda=$$532$~nm)
while the back-scattered light was recorded in the farfield by imaging the
Fourier plane of a microscope objective (oil immersion, NA=1.45).
The maximum observable momenta is overlaid as a coloured dashed circle.
All $L_{p}50-L_{p}30$ samples exhibit broad and statistically isotropic
scattering rings which resemble the primary resonance of the designed
structure factor shown in the inset of Fig.~\ref{FIG:KingsDesigns}D.
The momentum associated with the scattering resonance peak increases
with the downscaling of the sample; in $L_{p}25-L_{p}20$ the scattering
ring crosses over the observable momentum limit. The bright spot
at the centre of each farfield results from the specular reflection
along the axis of incidence.

Fig.~\ref{FIG:KingsDiffraction}C shows the azimuthally averaged
farfield intensity distributions. As the sample correlation length is reduced in size, the
momentum of the primary scattering peak increases with a small
intensity decrease. The structures $L_{p}50-L_{p}35$  are all characterised
by a single scattering peak with some finer structure
that varies from sample to sample. Interestingly, the $L_{p}30$ sample
exhibits a double peak, while $L_{p}25$ only a secondary low shoulder peak.
The linear scaling with the reciprocal of the correlation length,  shown in the insets of Figs.~\ref{FIG:KingsDiffraction}C,
can be expected by simple diffraction theory, while the finer structures are captured by
numerical finite-difference time-domain (FDTD) simulations shown in Fig.~\ref{FIG:KingsDiffraction}D.
In particular, FDTD predicts
the multi-peak signature of the $L_{p}30$ sample, suggesting that
this feature is a genuine property of the sample. We attribute this
finer structure to the interplay of the surface plasmon at the air/gold
interface sustained by a single pillar with the linear diffraction
dispersion.  In fact, the HuD structure comprises elements of different sizes,
much larger than the plasmon wavelength, which present high-order resonances in the visible range
and an overall response close to that of a surface plasmon resonance
of an infinite film which peaks around $k/k_{0}=1.09$.

\begin{figure}[!htb]
\centering
\includegraphics[width=7cm]{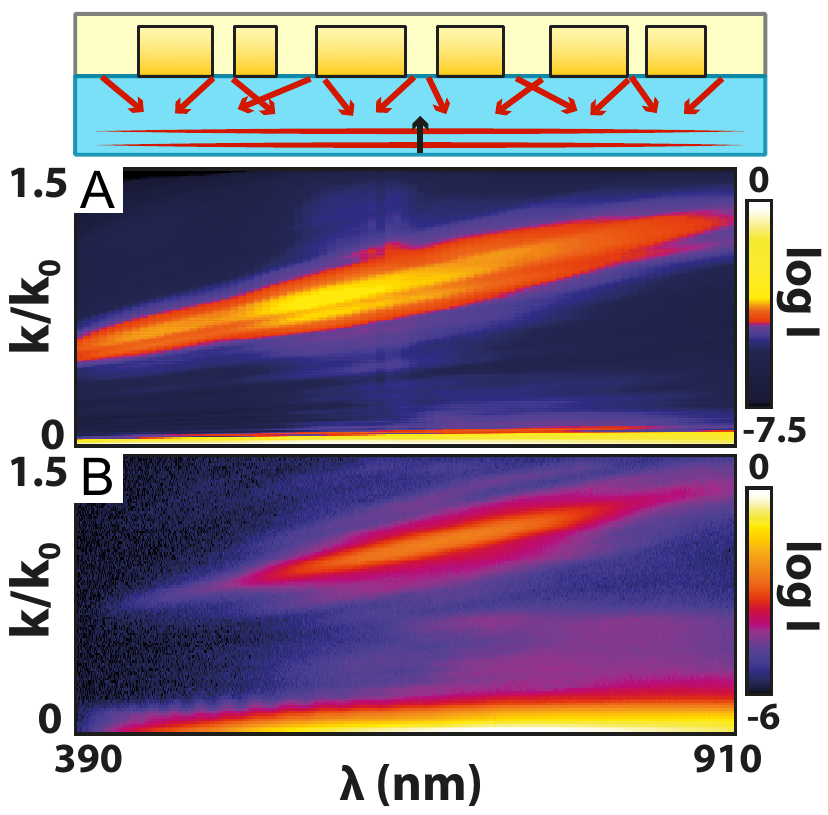}\caption{
Theoretical  (A) and experimental (B) white light scattering by the $L_{p}50$
metasurface. Both theory and experiment match well and show that the scattering peak momentum increases linearly with
wavelength. }
 \label{FIG:KingsWhiteLightScattering}
\end{figure}

We performed also a broadband scattering characterization of the $L_{p}50$
design. The sample was illuminated from below with white light, and
the reflected and scattered light was spectrally decomposed into its wavelength and
momentum components  (Fig.~\ref{FIG:KingsWhiteLightScattering}B)
by spectrally imaging the Fourier plane of the sample. In this way
an energy-wavevector dispersion diagram can be constructed as shown in Fig~\ref{FIG:KingsWhiteLightScattering}B
(experiments) and Fig~\ref{FIG:KingsWhiteLightScattering}A (FDTD calculations).
Both images display a bold diagonal slash, from low-scattering
angle to high-scattering angle, and the evolution, for increasing wavelengths, towards larger momenta of the primary scattering peak.  From this linear
relationship we conclude that the light diffraction follows the designed structure factor with a single main peak (Fig.~\ref{FIG:KingsDesigns}D).

\begin{figure}[htb]
\centering
\includegraphics[width=7cm]{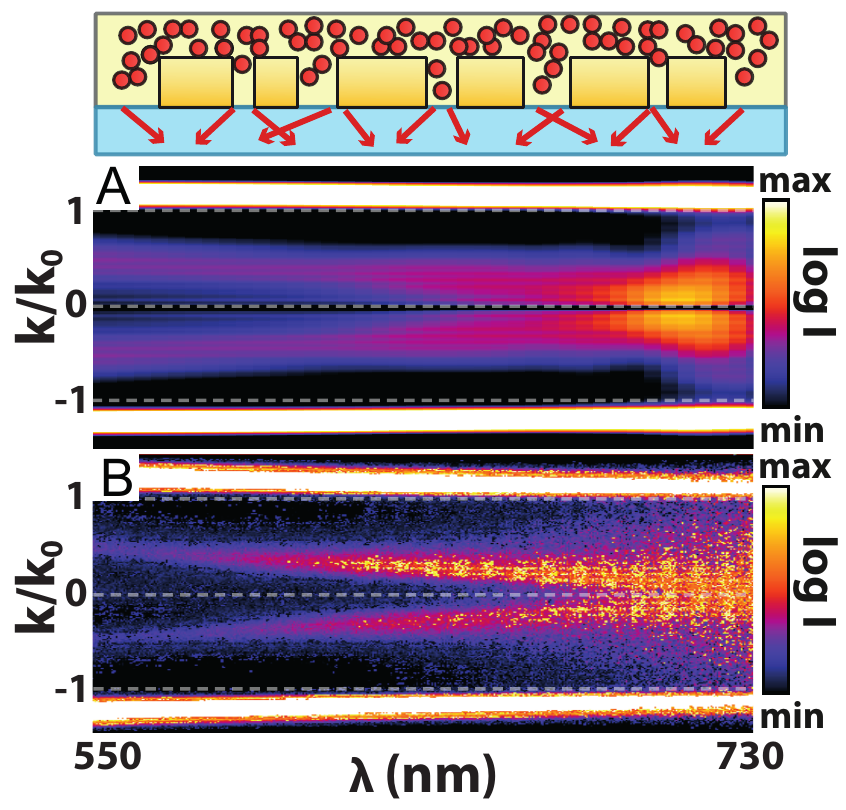}\caption{Generalised
dispersion relation $\omega(k)$ of $L_{p}50$ slab modes calculated summing
incoherently the farfields of 48 randomly oriented dipole (panel A) and measured by fluorescence
emission (panel B).}
 \label{FIG:KingsFluorescence}
\end{figure}

So far, we have investigated the ability of our gold metasurfaces
to mediate between incident and scattered light.  We
now  seek to characterise the HuD metasurface electromagnetic modes and their
momentum distribution. The metasurface modes are excited by a $50$~nm layer of Poly(methyl methacrylate) polymer highly doped with fluorescent 4-(Dicyanomethylene)-2-methyl-6-(4-dimethylaminostyryl)-4H-pyran (DCM)
dye molecules which was spin coated on the samples (Fig.~\ref{FIG:KingsFluorescence},
top panel).   Figs.~\ref{FIG:KingsFluorescence}A and B present
the theoretical and experimental frequency-momentum distribution, or dispersion plot, of the fluorescence
light emitted from the $L_{p}50$ structure when excited with a green
laser at a wavelength of 532 nm.

The pair of intense emission bands just outside the light lines (white dashed lines indicating $k/k_0 = \pm1$) results
from the characteristic radiation profile of a dipole near the glass-air
interface~\cite{Novotny2006}. Inside the light cone, we observe annular features
which describe the decomposition of the metasurface slab modes into
their in-plane momentum components. These may be viewed as the generalised
dispersion relation $\omega(\mathbf{k})$ of the slab modes.
Similar dispersion diagrams were measured for the $L_{p}50$, $L_{p}45$,
$L_{p}40$, $L_{p}35$ and $L_{p}30$ metasurfaces.

\begin{figure}[tbp]
\centering\includegraphics[width=7cm]{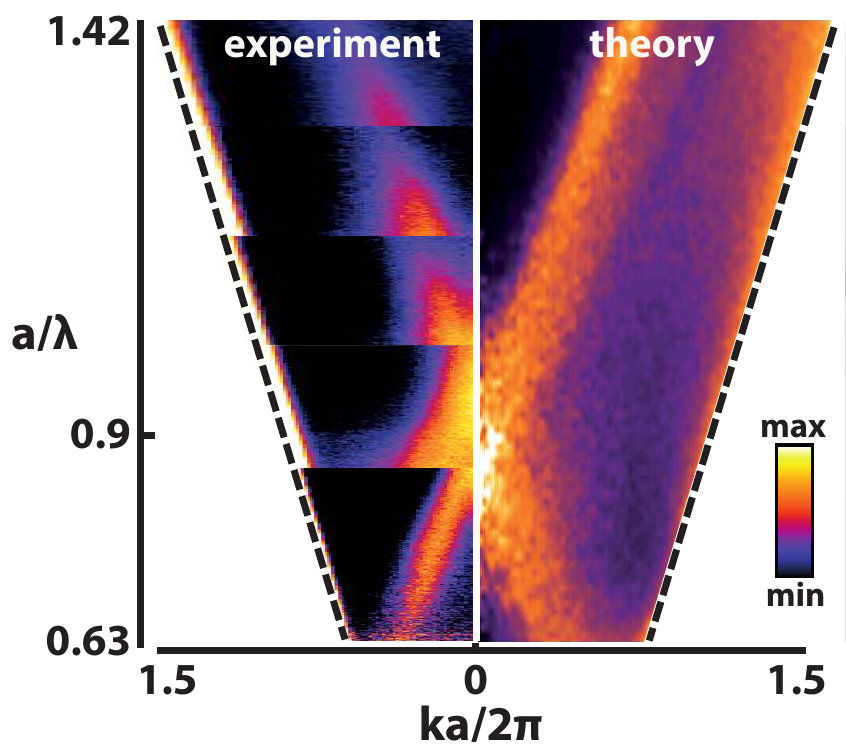}
\caption{Left, experimental dispersion relations
for samples $L_{p}50$, $L_{p}45$, $L_{p}40$, $L_{p}35$ and $L_{p}30$
are stacked together as a function of $a/\lambda$. Right panel, FDTD simulations
which confirm the crossing
at $a/\lambda$ = 0.9 where the light is emitted normal to the metasurface
plane.}
 \label{FIG:KingsSpectralFn}
\end{figure}

 In order to further analyse the dispersion plots of  Figs. \ref{FIG:KingsFluorescence},
  we convert the emission wavelength
and in-slab momentum to the dimensionless quantities $a/\lambda$
and $ka/2\pi$ respectively. The experimental dispersion diagrams
can then be stacked into a single image to illustrate the dispersion
over a large normalized frequency range (Fig.~\ref{FIG:KingsSpectralFn},
left panel).   It shows that the fluorescent light is emitted isotropically from
the HuD surface, into a cone with varying opening angle, which depends
on the emission wavelength to correlation length ratio.
As the frequency increases from $a/\lambda=0.63$, the dominant momentum
of the slab modes first decreases, reaching a zero in the region of
$a/\lambda=0.9$, at which point light is emitted normal to the slab
plane, before gradually opening back out. The experiments are in
good agreement with FDTD calculations (Fig.~\ref{FIG:KingsSpectralFn},
right panel) obtained with the spectral
function method ~\cite{Band2016}.
Specifically, decomposition of a slab's eigenmodes into a plane wave
basis can directly predict the farfield angular profile of its directional
emission. These results show that the electromagnetic dispersion diagram
of the HuD metasurface follows the designed structure factor (Figure \ref{FIG:KingsDesigns}D)
and exhibits band folding resulting from Bragg-like processes.

In conclusion, we designed, fabricated and characterized gold metasurfaces
derived from HuD connected networks. By reciprocal space
engineering we designed the structure factor to be dominated by a
single, broad scattering resonance which was observed to dictate
both the annular farfield light scattering and directional emission properties
of the metasurfaces. The observed HuD metasurface
dispersion corresponds to an effective
medium band that is back-folded by Bragg-like processes. Surface
plasmon resonance where found to contribute to the diffraction by
inducing additional peaks from their interaction with the structural diffraction peak.
The observed light emission and scattering engineering has important applications
for light extraction from light-emitting
devices, absorption in solar cells and annular redirection for displays, by exploiting the full azimuthal symmetry provided by the HuD metasurfaces design.

\section*{Acknowledgements}

The work was supported by the Engineering  and Physical Sciences Research Council (EPSRC EP/M027961/1 and EP/M013812/1 ),
the  Leverhulme  Trust (RPG-2014-238)  and  the  Royal  Society (RG140457).


\begin{thebibliography}{33}%
\makeatletter
\providecommand \@ifxundefined [1]{%
 \@ifx{#1\undefined}
}%
\providecommand \@ifnum [1]{%
 \ifnum #1\expandafter \@firstoftwo
 \else \expandafter \@secondoftwo
 \fi
}%
\providecommand \@ifx [1]{%
 \ifx #1\expandafter \@firstoftwo
 \else \expandafter \@secondoftwo
 \fi
}%
\providecommand \natexlab [1]{#1}%
\providecommand \enquote  [1]{``#1''}%
\providecommand \bibnamefont  [1]{#1}%
\providecommand \bibfnamefont [1]{#1}%
\providecommand \citenamefont [1]{#1}%
\providecommand \href@noop [0]{\@secondoftwo}%
\providecommand \href [0]{\begingroup \@sanitize@url \@href}%
\providecommand \@href[1]{\@@startlink{#1}\@@href}%
\providecommand \@@href[1]{\endgroup#1\@@endlink}%
\providecommand \@sanitize@url [0]{\catcode `\\12\catcode `\$12\catcode
  `\&12\catcode `\#12\catcode `\^12\catcode `\_12\catcode `\%12\relax}%
\providecommand \@@startlink[1]{}%
\providecommand \@@endlink[0]{}%
\providecommand \url  [0]{\begingroup\@sanitize@url \@url }%
\providecommand \@url [1]{\endgroup\@href {#1}{\urlprefix }}%
\providecommand \urlprefix  [0]{URL }%
\providecommand \Eprint [0]{\href }%
\providecommand \doibase [0]{http://dx.doi.org/}%
\providecommand \selectlanguage [0]{\@gobble}%
\providecommand \bibinfo  [0]{\@secondoftwo}%
\providecommand \bibfield  [0]{\@secondoftwo}%
\providecommand \translation [1]{[#1]}%
\providecommand \BibitemOpen [0]{}%
\providecommand \bibitemStop [0]{}%
\providecommand \bibitemNoStop [0]{.\EOS\space}%
\providecommand \EOS [0]{\spacefactor3000\relax}%
\providecommand \BibitemShut  [1]{\csname bibitem#1\endcsname}%
\let\auto@bib@innerbib\@empty
\bibitem [{\citenamefont {Gholipour}\ \emph {et~al.}(2013)\citenamefont
  {Gholipour}, \citenamefont {Zhang}, \citenamefont {MacDonald}, \citenamefont
  {Hewak},\ and\ \citenamefont {Zheludev}}]{Gholipour2013}%
  \BibitemOpen
  \bibfield  {author} {\bibinfo {author} {\bibfnamefont {Behrad}\ \bibnamefont
  {Gholipour}}, \bibinfo {author} {\bibfnamefont {Jianfa}\ \bibnamefont
  {Zhang}}, \bibinfo {author} {\bibfnamefont {Kevin~F.}\ \bibnamefont
  {MacDonald}}, \bibinfo {author} {\bibfnamefont {Daniel~W.}\ \bibnamefont
  {Hewak}}, \ and\ \bibinfo {author} {\bibfnamefont {Nikolay~I.}\ \bibnamefont
  {Zheludev}},\ }\bibfield  {title} {\enquote {\bibinfo {title} {{An
  all-optical, non-volatile, bidirectional, phase-change meta-switch}},}\
  }\href {\doibase 10.1002/adma.201300588} {\bibfield  {journal} {\bibinfo
  {journal} {Adv. Mater.}\ }\textbf {\bibinfo {volume} {25}},\ \bibinfo {pages}
  {3050--3054} (\bibinfo {year} {2013})}\BibitemShut {NoStop}%
\bibitem [{\citenamefont {Hosseini}\ \emph {et~al.}(2014)\citenamefont
  {Hosseini}, \citenamefont {Wright},\ and\ \citenamefont
  {Bhaskaran}}]{Hosseini2014}%
  \BibitemOpen
  \bibfield  {author} {\bibinfo {author} {\bibfnamefont {P}~\bibnamefont
  {Hosseini}}, \bibinfo {author} {\bibfnamefont {C~D}\ \bibnamefont {Wright}},
  \ and\ \bibinfo {author} {\bibfnamefont {H}~\bibnamefont {Bhaskaran}},\
  }\bibfield  {title} {\enquote {\bibinfo {title} {{An optoelectronic framework
  enabled by low-dimensional phase-change films}},}\ }\href {\doibase
  10.1038/nature13487} {\bibfield  {journal} {\bibinfo  {journal} {Nature}\
  }\textbf {\bibinfo {volume} {511}},\ \bibinfo {pages} {206--211} (\bibinfo
  {year} {2014})}\BibitemShut {NoStop}%
\bibitem [{\citenamefont {Azad}\ \emph {et~al.}(2016)\citenamefont {Azad},
  \citenamefont {Kort-Kamp}, \citenamefont {Sykora}, \citenamefont
  {Weisse-Bernstein}, \citenamefont {Luk}, \citenamefont {Taylor},
  \citenamefont {Dalvit},\ and\ \citenamefont {Chen}}]{Azad2016}%
  \BibitemOpen
  \bibfield  {author} {\bibinfo {author} {\bibfnamefont {Abul~K}\ \bibnamefont
  {Azad}}, \bibinfo {author} {\bibfnamefont {Wilton J~M}\ \bibnamefont
  {Kort-Kamp}}, \bibinfo {author} {\bibfnamefont {Milan}\ \bibnamefont
  {Sykora}}, \bibinfo {author} {\bibfnamefont {Nina~R}\ \bibnamefont
  {Weisse-Bernstein}}, \bibinfo {author} {\bibfnamefont {Ting~S}\ \bibnamefont
  {Luk}}, \bibinfo {author} {\bibfnamefont {Antoinette~J}\ \bibnamefont
  {Taylor}}, \bibinfo {author} {\bibfnamefont {Diego A~R}\ \bibnamefont
  {Dalvit}}, \ and\ \bibinfo {author} {\bibfnamefont {Hou-Tong}\ \bibnamefont
  {Chen}},\ }\bibfield  {title} {\enquote {\bibinfo {title} {{Metasurface
  Broadband Solar Absorber.}}}\ }\href {\doibase 10.1038/srep20347} {\bibfield
  {journal} {\bibinfo  {journal} {Sci. Rep.}\ }\textbf {\bibinfo {volume}
  {6}},\ \bibinfo {pages} {20347} (\bibinfo {year} {2016})}\BibitemShut
  {NoStop}%
\bibitem [{\citenamefont {Almoneef}\ and\ \citenamefont
  {Ramahi}(2015)}]{Almoneef2015}%
  \BibitemOpen
  \bibfield  {author} {\bibinfo {author} {\bibfnamefont {Thamer~S.}\
  \bibnamefont {Almoneef}}\ and\ \bibinfo {author} {\bibfnamefont {Omar~M.}\
  \bibnamefont {Ramahi}},\ }\bibfield  {title} {\enquote {\bibinfo {title}
  {{Metamaterial electromagnetic energy harvester with near unity
  efficiency}},}\ }\href {\doibase 10.1063/1.4916232} {\bibfield  {journal}
  {\bibinfo  {journal} {Appl. Phys. Lett.}\ }\textbf {\bibinfo {volume} {106}}
  (\bibinfo {year} {2015}),\ 10.1063/1.4916232}\BibitemShut {NoStop}%
\bibitem [{\citenamefont {Barnes}\ \emph {et~al.}(1995)\citenamefont {Barnes},
  \citenamefont {Preist}, \citenamefont {Kitson}, \citenamefont {Sambles},
  \citenamefont {Cotter},\ and\ \citenamefont {Nash}}]{Barnes1995}%
  \BibitemOpen
  \bibfield  {author} {\bibinfo {author} {\bibfnamefont {W.~L.}\ \bibnamefont
  {Barnes}}, \bibinfo {author} {\bibfnamefont {T.~W.}\ \bibnamefont {Preist}},
  \bibinfo {author} {\bibfnamefont {S.~C.}\ \bibnamefont {Kitson}}, \bibinfo
  {author} {\bibfnamefont {J.~R.}\ \bibnamefont {Sambles}}, \bibinfo {author}
  {\bibfnamefont {N.~P~K}\ \bibnamefont {Cotter}}, \ and\ \bibinfo {author}
  {\bibfnamefont {D.~J.}\ \bibnamefont {Nash}},\ }\bibfield  {title} {\enquote
  {\bibinfo {title} {{Photonic gaps in the dispersion of surface plasmons on
  gratings}},}\ }\href {\doibase 10.1103/PhysRevB.51.11164} {\bibfield
  {journal} {\bibinfo  {journal} {Phys. Rev. B}\ }\textbf {\bibinfo {volume}
  {51}},\ \bibinfo {pages} {11164--11167} (\bibinfo {year} {1995})}\BibitemShut
  {NoStop}%
\bibitem [{\citenamefont {Bouillard}\ \emph {et~al.}(2014)\citenamefont
  {Bouillard}, \citenamefont {Segovia}, \citenamefont {Dickson}, \citenamefont
  {Wurtz},\ and\ \citenamefont {Zayats}}]{Bouillard2014}%
  \BibitemOpen
  \bibfield  {author} {\bibinfo {author} {\bibfnamefont {J-S}\ \bibnamefont
  {Bouillard}}, \bibinfo {author} {\bibfnamefont {P}~\bibnamefont {Segovia}},
  \bibinfo {author} {\bibfnamefont {W}~\bibnamefont {Dickson}}, \bibinfo
  {author} {\bibfnamefont {G~a}\ \bibnamefont {Wurtz}}, \ and\ \bibinfo
  {author} {\bibfnamefont {a~V}\ \bibnamefont {Zayats}},\ }\bibfield  {title}
  {\enquote {\bibinfo {title} {{Shaping plasmon beams via the controlled
  illumination of finite-size plasmonic crystals.}}}\ }\href {\doibase
  10.1038/srep07234} {\bibfield  {journal} {\bibinfo  {journal} {Sci. Rep.}\
  }\textbf {\bibinfo {volume} {4}},\ \bibinfo {pages} {7234} (\bibinfo {year}
  {2014})}\BibitemShut {NoStop}%
\bibitem [{\citenamefont {Giannini}\ \emph {et~al.}(2010)\citenamefont
  {Giannini}, \citenamefont {Vecchi},\ and\ \citenamefont
  {G\'omez~Rivas}}]{Giannini2010}%
  \BibitemOpen
  \bibfield  {author} {\bibinfo {author} {\bibfnamefont {V.}~\bibnamefont
  {Giannini}}, \bibinfo {author} {\bibfnamefont {G.}~\bibnamefont {Vecchi}}, \
  and\ \bibinfo {author} {\bibfnamefont {J.}~\bibnamefont {G\'omez~Rivas}},\
  }\bibfield  {title} {\enquote {\bibinfo {title} {Lighting up multipolar
  surface plasmon polaritons by collective resonances in arrays of
  nanoantennas},}\ }\href {\doibase 10.1103/PhysRevLett.105.266801} {\bibfield
  {journal} {\bibinfo  {journal} {Phys. Rev. Lett.}\ }\textbf {\bibinfo
  {volume} {105}},\ \bibinfo {pages} {266801} (\bibinfo {year}
  {2010})}\BibitemShut {NoStop}%
\bibitem [{\citenamefont {{Dal Negro}}\ and\ \citenamefont
  {Boriskina}(2012)}]{DalNegro2012}%
  \BibitemOpen
  \bibfield  {author} {\bibinfo {author} {\bibfnamefont {L.}~\bibnamefont {{Dal
  Negro}}}\ and\ \bibinfo {author} {\bibfnamefont {S.~V.}\ \bibnamefont
  {Boriskina}},\ }\bibfield  {title} {\enquote {\bibinfo {title}
  {{Deterministic aperiodic nanostructures for photonics and plasmonics
  applications}},}\ }\href {\doibase 10.1002/lpor.201000046} {\bibfield
  {journal} {\bibinfo  {journal} {Laser Photonics Rev.}\ }\textbf {\bibinfo
  {volume} {6}},\ \bibinfo {pages} {178--218} (\bibinfo {year}
  {2012})}\BibitemShut {NoStop}%
\bibitem [{\citenamefont {Lubin}\ \emph {et~al.}(2013)\citenamefont {Lubin},
  \citenamefont {Hryn}, \citenamefont {Huntington}, \citenamefont {Engel},\
  and\ \citenamefont {Odom}}]{Lubin2013}%
  \BibitemOpen
  \bibfield  {author} {\bibinfo {author} {\bibfnamefont {Steven~M.}\
  \bibnamefont {Lubin}}, \bibinfo {author} {\bibfnamefont {Alexander~J.}\
  \bibnamefont {Hryn}}, \bibinfo {author} {\bibfnamefont {Mark~D.}\
  \bibnamefont {Huntington}}, \bibinfo {author} {\bibfnamefont {Clifford~J.}\
  \bibnamefont {Engel}}, \ and\ \bibinfo {author} {\bibfnamefont {Teri~W.}\
  \bibnamefont {Odom}},\ }\bibfield  {title} {\enquote {\bibinfo {title}
  {{Quasiperiodic moir{\'{e}} plasmonic crystals}},}\ }\href {\doibase
  10.1021/nn404703z} {\bibfield  {journal} {\bibinfo  {journal} {ACS Nano}\
  }\textbf {\bibinfo {volume} {7}},\ \bibinfo {pages} {11035--11042} (\bibinfo
  {year} {2013})}\BibitemShut {NoStop}%
\bibitem [{\citenamefont {Lawrence}\ \emph {et~al.}(2012)\citenamefont
  {Lawrence}, \citenamefont {Trevino},\ and\ \citenamefont {{Dal
  Negro}}}]{Lawrence2012}%
  \BibitemOpen
  \bibfield  {author} {\bibinfo {author} {\bibfnamefont {Nate}\ \bibnamefont
  {Lawrence}}, \bibinfo {author} {\bibfnamefont {Jacob}\ \bibnamefont
  {Trevino}}, \ and\ \bibinfo {author} {\bibfnamefont {Luca}\ \bibnamefont
  {{Dal Negro}}},\ }\bibfield  {title} {\enquote {\bibinfo {title} {{Aperiodic
  arrays of active nanopillars for radiation engineering}},}\ }\href {\doibase
  10.1063/1.4723564} {\bibfield  {journal} {\bibinfo  {journal} {J. Appl.
  Phys.}\ }\textbf {\bibinfo {volume} {111}} (\bibinfo {year} {2012}),\
  10.1063/1.4723564}\BibitemShut {NoStop}%
\bibitem [{\citenamefont {Afshinmanesh}\ \emph {et~al.}(2014)\citenamefont
  {Afshinmanesh}, \citenamefont {Curto}, \citenamefont {Milaninia},
  \citenamefont {{Van Hulst}},\ and\ \citenamefont
  {Brongersma}}]{Afshinmanesh2014}%
  \BibitemOpen
  \bibfield  {author} {\bibinfo {author} {\bibfnamefont {Farzaneh}\
  \bibnamefont {Afshinmanesh}}, \bibinfo {author} {\bibfnamefont {Alberto~G.}\
  \bibnamefont {Curto}}, \bibinfo {author} {\bibfnamefont {Kaveh~M.}\
  \bibnamefont {Milaninia}}, \bibinfo {author} {\bibfnamefont {Niek~F.}\
  \bibnamefont {{Van Hulst}}}, \ and\ \bibinfo {author} {\bibfnamefont
  {Mark~L.}\ \bibnamefont {Brongersma}},\ }\bibfield  {title} {\enquote
  {\bibinfo {title} {{Transparent metallic fractal electrodes for semiconductor
  devices}},}\ }\href {\doibase 10.1021/nl501738b} {\bibfield  {journal}
  {\bibinfo  {journal} {Nano Lett.}\ }\textbf {\bibinfo {volume} {14}},\
  \bibinfo {pages} {5068--5074} (\bibinfo {year} {2014})}\BibitemShut {NoStop}%
\bibitem [{\citenamefont {Gaio}\ \emph {et~al.}(2015)\citenamefont {Gaio},
  \citenamefont {Castro-Lopez}, \citenamefont {Renger}, \citenamefont {van
  Hulst},\ and\ \citenamefont {Sapienza}}]{Gaio2015a}%
  \BibitemOpen
  \bibfield  {author} {\bibinfo {author} {\bibfnamefont {Michele}\ \bibnamefont
  {Gaio}}, \bibinfo {author} {\bibfnamefont {Marta}\ \bibnamefont
  {Castro-Lopez}}, \bibinfo {author} {\bibfnamefont {Jan}\ \bibnamefont
  {Renger}}, \bibinfo {author} {\bibfnamefont {Niek}\ \bibnamefont {van
  Hulst}}, \ and\ \bibinfo {author} {\bibfnamefont {Riccardo}\ \bibnamefont
  {Sapienza}},\ }\bibfield  {title} {\enquote {\bibinfo {title} {{Percolating
  plasmonic networks for light emission control}},}\ }\href {\doibase
  10.1039/C4FD00187G} {\bibfield  {journal} {\bibinfo  {journal} {Faraday
  Discuss.}\ }\textbf {\bibinfo {volume} {178}},\ \bibinfo {pages} {237--252}
  (\bibinfo {year} {2015})}\BibitemShut {NoStop}%
\bibitem [{\citenamefont {Schokker}\ and\ \citenamefont
  {Koenderink}(2015)}]{Schokker2015}%
  \BibitemOpen
  \bibfield  {author} {\bibinfo {author} {\bibfnamefont {A.~Hinke}\
  \bibnamefont {Schokker}}\ and\ \bibinfo {author} {\bibfnamefont {A.~Femius}\
  \bibnamefont {Koenderink}},\ }\bibfield  {title} {\enquote {\bibinfo {title}
  {Statistics of randomized plasmonic lattice lasers},}\ }\href {\doibase
  10.1021/acsphotonics.5b00226} {\bibfield  {journal} {\bibinfo  {journal} {ACS
  Photonics}\ }\textbf {\bibinfo {volume} {2}},\ \bibinfo {pages} {1289--1297}
  (\bibinfo {year} {2015})}\BibitemShut {NoStop}%
\bibitem [{\citenamefont {Zhang}\ \emph {et~al.}(2016)\citenamefont {Zhang},
  \citenamefont {Knitter}, \citenamefont {Liew}, \citenamefont {Omenetto},
  \citenamefont {Reinhard}, \citenamefont {Cao},\ and\ \citenamefont {{Dal
  Negro}}}]{Zhang2016}%
  \BibitemOpen
  \bibfield  {author} {\bibinfo {author} {\bibfnamefont {R.}~\bibnamefont
  {Zhang}}, \bibinfo {author} {\bibfnamefont {S.}~\bibnamefont {Knitter}},
  \bibinfo {author} {\bibfnamefont {S.~F.}\ \bibnamefont {Liew}}, \bibinfo
  {author} {\bibfnamefont {F.~G.}\ \bibnamefont {Omenetto}}, \bibinfo {author}
  {\bibfnamefont {B.~M.}\ \bibnamefont {Reinhard}}, \bibinfo {author}
  {\bibfnamefont {H.}~\bibnamefont {Cao}}, \ and\ \bibinfo {author}
  {\bibfnamefont {L.}~\bibnamefont {{Dal Negro}}},\ }\bibfield  {title}
  {\enquote {\bibinfo {title} {{Plasmon-enhanced random lasing in
  bio-compatible networks of cellulose nanofibers}},}\ }\href {\doibase
  10.1063/1.4939263} {\bibfield  {journal} {\bibinfo  {journal} {Appl. Phys.
  Lett.}\ }\textbf {\bibinfo {volume} {108}} (\bibinfo {year} {2016}),\
  10.1063/1.4939263}\BibitemShut {NoStop}%
\bibitem [{\citenamefont {Yu}\ \emph {et~al.}(2011)\citenamefont {Yu},
  \citenamefont {Genevet}, \citenamefont {Kats}, \citenamefont {Aieta},
  \citenamefont {Tetienne}, \citenamefont {Capasso},\ and\ \citenamefont
  {Gaburro}}]{Capasso2011}%
  \BibitemOpen
  \bibfield  {author} {\bibinfo {author} {\bibfnamefont {Nanfang}\ \bibnamefont
  {Yu}}, \bibinfo {author} {\bibfnamefont {Patrice}\ \bibnamefont {Genevet}},
  \bibinfo {author} {\bibfnamefont {Mikhail~A}\ \bibnamefont {Kats}}, \bibinfo
  {author} {\bibfnamefont {Francesco}\ \bibnamefont {Aieta}}, \bibinfo {author}
  {\bibfnamefont {Jean-Philippe}\ \bibnamefont {Tetienne}}, \bibinfo {author}
  {\bibfnamefont {Federico}\ \bibnamefont {Capasso}}, \ and\ \bibinfo {author}
  {\bibfnamefont {Zeno}\ \bibnamefont {Gaburro}},\ }\bibfield  {title}
  {\enquote {\bibinfo {title} {{Light propagation with phase discontinuities:
  generalized laws of reflection and refraction.}}}\ }\href {\doibase
  10.1126/science.1210713} {\bibfield  {journal} {\bibinfo  {journal}
  {Science}\ }\textbf {\bibinfo {volume} {334}},\ \bibinfo {pages} {333--7}
  (\bibinfo {year} {2011})}\BibitemShut {NoStop}%
\bibitem [{\citenamefont {Ni}\ \emph {et~al.}(2013)\citenamefont {Ni},
  \citenamefont {Kildishev},\ and\ \citenamefont {Shalaev}}]{Ni2013}%
  \BibitemOpen
  \bibfield  {author} {\bibinfo {author} {\bibfnamefont {Xingjie}\ \bibnamefont
  {Ni}}, \bibinfo {author} {\bibfnamefont {Alexander~V}\ \bibnamefont
  {Kildishev}}, \ and\ \bibinfo {author} {\bibfnamefont {Vladimir~M}\
  \bibnamefont {Shalaev}},\ }\bibfield  {title} {\enquote {\bibinfo {title}
  {{Metasurface holograms for visible light}},}\ }\href {\doibase
  10.1038/ncomms3807} {\bibfield  {journal} {\bibinfo  {journal} {Nat.
  Commun.}\ }\textbf {\bibinfo {volume} {4}},\ \bibinfo {pages} {2807}
  (\bibinfo {year} {2013})}\BibitemShut {NoStop}%
\bibitem [{\citenamefont {Martins}\ \emph {et~al.}(2013)\citenamefont
  {Martins}, \citenamefont {Li}, \citenamefont {Liu}, \citenamefont {Depauw},
  \citenamefont {Chen}, \citenamefont {Zhou},\ and\ \citenamefont
  {Krauss}}]{Martins2013}%
  \BibitemOpen
  \bibfield  {author} {\bibinfo {author} {\bibfnamefont {Emiliano~R}\
  \bibnamefont {Martins}}, \bibinfo {author} {\bibfnamefont {Juntao}\
  \bibnamefont {Li}}, \bibinfo {author} {\bibfnamefont {YiKun}\ \bibnamefont
  {Liu}}, \bibinfo {author} {\bibfnamefont {Val{\'{e}}rie}\ \bibnamefont
  {Depauw}}, \bibinfo {author} {\bibfnamefont {Zhanxu}\ \bibnamefont {Chen}},
  \bibinfo {author} {\bibfnamefont {Jianying}\ \bibnamefont {Zhou}}, \ and\
  \bibinfo {author} {\bibfnamefont {Thomas~F}\ \bibnamefont {Krauss}},\
  }\bibfield  {title} {\enquote {\bibinfo {title} {{Deterministic quasi-random
  nanostructures for photon control.}}}\ }\href {\doibase 10.1038/ncomms3665}
  {\bibfield  {journal} {\bibinfo  {journal} {Nat. Commun.}\ }\textbf {\bibinfo
  {volume} {4}},\ \bibinfo {pages} {2665} (\bibinfo {year} {2013})}\BibitemShut
  {NoStop}%
\bibitem [{\citenamefont {Burresi}\ \emph {et~al.}(2013)\citenamefont
  {Burresi}, \citenamefont {Pratesi}, \citenamefont {Vynck}, \citenamefont
  {Prasciolu}, \citenamefont {Tormen},\ and\ \citenamefont
  {Wiersma}}]{Burresi2013}%
  \BibitemOpen
  \bibfield  {author} {\bibinfo {author} {\bibfnamefont {Matteo}\ \bibnamefont
  {Burresi}}, \bibinfo {author} {\bibfnamefont {Filippo}\ \bibnamefont
  {Pratesi}}, \bibinfo {author} {\bibfnamefont {Kevin}\ \bibnamefont {Vynck}},
  \bibinfo {author} {\bibfnamefont {Mauro}\ \bibnamefont {Prasciolu}}, \bibinfo
  {author} {\bibfnamefont {Massimo}\ \bibnamefont {Tormen}}, \ and\ \bibinfo
  {author} {\bibfnamefont {Diederik~S.}\ \bibnamefont {Wiersma}},\ }\bibfield
  {title} {\enquote {\bibinfo {title} {Two-dimensional disorder for broadband,
  omnidirectional and polarization-insensitive absorption},}\ }\href {\doibase
  10.1364/OE.21.00A268} {\bibfield  {journal} {\bibinfo  {journal} {Opt.
  Express}\ }\textbf {\bibinfo {volume} {21}},\ \bibinfo {pages} {A268--A275}
  (\bibinfo {year} {2013})}\BibitemShut {NoStop}%
\bibitem [{\citenamefont {Lagendijk}\ \emph {et~al.}(2009)\citenamefont
  {Lagendijk}, \citenamefont {{Van Tiggelen}},\ and\ \citenamefont
  {Wiersma}}]{Lagendijk2009}%
  \BibitemOpen
  \bibfield  {author} {\bibinfo {author} {\bibfnamefont {Ad}~\bibnamefont
  {Lagendijk}}, \bibinfo {author} {\bibfnamefont {Bart}\ \bibnamefont {{Van
  Tiggelen}}}, \ and\ \bibinfo {author} {\bibfnamefont {Diederik~S.}\
  \bibnamefont {Wiersma}},\ }\bibfield  {title} {\enquote {\bibinfo {title}
  {{Fifty years of Anderson localization}},}\ }\href {\doibase
  10.1063/1.3206091} {\bibfield  {journal} {\bibinfo  {journal} {Phys. Today}\
  }\textbf {\bibinfo {volume} {62}},\ \bibinfo {pages} {24--29} (\bibinfo
  {year} {2009})}\BibitemShut {NoStop}%
\bibitem [{\citenamefont {Koo}\ \emph {et~al.}(2010)\citenamefont {Koo},
  \citenamefont {Jeong}, \citenamefont {Araoka}, \citenamefont {Ishikawa},
  \citenamefont {Nishimura}, \citenamefont {Toyooka},\ and\ \citenamefont
  {Takezoe}}]{Koo2010}%
  \BibitemOpen
  \bibfield  {author} {\bibinfo {author} {\bibfnamefont {Won~Hoe}\ \bibnamefont
  {Koo}}, \bibinfo {author} {\bibfnamefont {Soon~Moon}\ \bibnamefont {Jeong}},
  \bibinfo {author} {\bibfnamefont {Fumito}\ \bibnamefont {Araoka}}, \bibinfo
  {author} {\bibfnamefont {Ken}\ \bibnamefont {Ishikawa}}, \bibinfo {author}
  {\bibfnamefont {Suzushi}\ \bibnamefont {Nishimura}}, \bibinfo {author}
  {\bibfnamefont {Takehiro}\ \bibnamefont {Toyooka}}, \ and\ \bibinfo {author}
  {\bibfnamefont {Hideo}\ \bibnamefont {Takezoe}},\ }\bibfield  {title}
  {\enquote {\bibinfo {title} {{Light extraction from organic light-emitting
  diodes enhanced by spontaneously formed buckles}},}\ }\href {\doibase
  10.1038/nphoton.2010.7} {\bibfield  {journal} {\bibinfo  {journal} {Nat.
  Photonics}\ }\textbf {\bibinfo {volume} {4}},\ \bibinfo {pages} {222--226}
  (\bibinfo {year} {2010})}\BibitemShut {NoStop}%
\bibitem [{\citenamefont {Froufe-P\'erez}\ \emph {et~al.}(2016)\citenamefont
  {Froufe-P\'erez}, \citenamefont {Engel}, \citenamefont {Damasceno},
  \citenamefont {Muller}, \citenamefont {Haberko}, \citenamefont {Glotzer},\
  and\ \citenamefont {Scheffold}}]{FroufePRL}%
  \BibitemOpen
  \bibfield  {author} {\bibinfo {author} {\bibfnamefont {Luis~S.}\ \bibnamefont
  {Froufe-P\'erez}}, \bibinfo {author} {\bibfnamefont {Michael}\ \bibnamefont
  {Engel}}, \bibinfo {author} {\bibfnamefont {Pablo~F.}\ \bibnamefont
  {Damasceno}}, \bibinfo {author} {\bibfnamefont {Nicolas}\ \bibnamefont
  {Muller}}, \bibinfo {author} {\bibfnamefont {Jakub}\ \bibnamefont {Haberko}},
  \bibinfo {author} {\bibfnamefont {Sharon~C.}\ \bibnamefont {Glotzer}}, \ and\
  \bibinfo {author} {\bibfnamefont {Frank}\ \bibnamefont {Scheffold}},\
  }\bibfield  {title} {\enquote {\bibinfo {title} {Role of short-range order
  and hyperuniformity in the formation of band gaps in disordered photonic
  materials},}\ }\href {\doibase 10.1103/PhysRevLett.117.053902} {\bibfield
  {journal} {\bibinfo  {journal} {Phys. Rev. Lett.}\ }\textbf {\bibinfo
  {volume} {117}},\ \bibinfo {pages} {053902} (\bibinfo {year}
  {2016})}\BibitemShut {NoStop}%
\bibitem [{\citenamefont {Florescu}\ \emph
  {et~al.}(2009{\natexlab{a}})\citenamefont {Florescu}, \citenamefont
  {Torquato},\ and\ \citenamefont {Steinhardt}}]{Florescu2009}%
  \BibitemOpen
  \bibfield  {author} {\bibinfo {author} {\bibfnamefont {Marian}\ \bibnamefont
  {Florescu}}, \bibinfo {author} {\bibfnamefont {Salvatore}\ \bibnamefont
  {Torquato}}, \ and\ \bibinfo {author} {\bibfnamefont {Paul~J}\ \bibnamefont
  {Steinhardt}},\ }\bibfield  {title} {\enquote {\bibinfo {title} {{Designer
  disordered materials with large, complete photonic band gaps}},}\ }\href
  {\doibase 10.1073/pnas.0907744106} {\bibfield  {journal} {\bibinfo  {journal}
  {Proc. Natl. Acad. Sci.}\ }\textbf {\bibinfo {volume} {106}},\ \bibinfo
  {pages} {20658--20663} (\bibinfo {year} {2009}{\natexlab{a}})}\BibitemShut
  {NoStop}%
\bibitem [{\citenamefont {Florescu}\ \emph
  {et~al.}(2009{\natexlab{b}})\citenamefont {Florescu}, \citenamefont
  {Torquato},\ and\ \citenamefont {Steinhardt}}]{Florescu2009b}%
  \BibitemOpen
  \bibfield  {author} {\bibinfo {author} {\bibfnamefont {Marian}\ \bibnamefont
  {Florescu}}, \bibinfo {author} {\bibfnamefont {Salvatore}\ \bibnamefont
  {Torquato}}, \ and\ \bibinfo {author} {\bibfnamefont {Paul~J.}\ \bibnamefont
  {Steinhardt}},\ }\bibfield  {title} {\enquote {\bibinfo {title} {{Complete
  band gaps in two-dimensional photonic quasicrystals}},}\ }\href {\doibase
  10.1103/PhysRevB.80.155112} {\bibfield  {journal} {\bibinfo  {journal} {Phys.
  Rev. B - Condens. Matter Mater. Phys.}\ }\textbf {\bibinfo {volume} {80}}
  (\bibinfo {year} {2009}{\natexlab{b}}),\
  10.1103/PhysRevB.80.155112}\BibitemShut {NoStop}%
\bibitem [{\citenamefont {Muller}\ \emph {et~al.}(2014)\citenamefont {Muller},
  \citenamefont {Haberko}, \citenamefont {Marichy},\ and\ \citenamefont
  {Scheffold}}]{Muller2014}%
  \BibitemOpen
  \bibfield  {author} {\bibinfo {author} {\bibfnamefont {Nicolas}\ \bibnamefont
  {Muller}}, \bibinfo {author} {\bibfnamefont {Jakub}\ \bibnamefont {Haberko}},
  \bibinfo {author} {\bibfnamefont {Catherine}\ \bibnamefont {Marichy}}, \ and\
  \bibinfo {author} {\bibfnamefont {Frank}\ \bibnamefont {Scheffold}},\
  }\bibfield  {title} {\enquote {\bibinfo {title} {{Silicon hyperuniform
  disordered photonic materials with a pronounced gap in the shortwave
  infrared}},}\ }\href {\doibase 10.1002/adom.201300415} {\bibfield  {journal}
  {\bibinfo  {journal} {Adv. Opt. Mater.}\ }\textbf {\bibinfo {volume} {2}},\
  \bibinfo {pages} {115--119} (\bibinfo {year} {2014})}\BibitemShut {NoStop}%
\bibitem [{\citenamefont {Man}\ \emph {et~al.}(2013)\citenamefont {Man},
  \citenamefont {Florescu}, \citenamefont {Williamson}, \citenamefont {He},
  \citenamefont {Hashemizad}, \citenamefont {Leung}, \citenamefont {Liner},
  \citenamefont {Torquato}, \citenamefont {Chaikin},\ and\ \citenamefont
  {Steinhardt}}]{Man2013}%
  \BibitemOpen
  \bibfield  {author} {\bibinfo {author} {\bibfnamefont {Weining}\ \bibnamefont
  {Man}}, \bibinfo {author} {\bibfnamefont {Marian}\ \bibnamefont {Florescu}},
  \bibinfo {author} {\bibfnamefont {Eric~Paul}\ \bibnamefont {Williamson}},
  \bibinfo {author} {\bibfnamefont {Yingquan}\ \bibnamefont {He}}, \bibinfo
  {author} {\bibfnamefont {Seyed~Reza}\ \bibnamefont {Hashemizad}}, \bibinfo
  {author} {\bibfnamefont {Brian Y~C}\ \bibnamefont {Leung}}, \bibinfo {author}
  {\bibfnamefont {Devin~Robert}\ \bibnamefont {Liner}}, \bibinfo {author}
  {\bibfnamefont {Salvatore}\ \bibnamefont {Torquato}}, \bibinfo {author}
  {\bibfnamefont {Paul~M}\ \bibnamefont {Chaikin}}, \ and\ \bibinfo {author}
  {\bibfnamefont {Paul~J}\ \bibnamefont {Steinhardt}},\ }\bibfield  {title}
  {\enquote {\bibinfo {title} {{Isotropic band gaps and freeform waveguides
  observed in hyperuniform disordered photonic solids}},}\ }\href {\doibase
  10.1073/pnas.1307879110} {\bibfield  {journal} {\bibinfo  {journal} {Proc.
  Natl. Acad. Sci.}\ }\textbf {\bibinfo {volume} {110}},\ \bibinfo {pages}
  {15886--15891} (\bibinfo {year} {2013})}\BibitemShut {NoStop}%
\bibitem [{\citenamefont {Zhou}\ \emph {et~al.}(2016)\citenamefont {Zhou},
  \citenamefont {Cheng}, \citenamefont {Zhu}, \citenamefont {Sun},\ and\
  \citenamefont {Tsang}}]{Zhou2016}%
  \BibitemOpen
  \bibfield  {author} {\bibinfo {author} {\bibfnamefont {Wen}\ \bibnamefont
  {Zhou}}, \bibinfo {author} {\bibfnamefont {Zhenzhou}\ \bibnamefont {Cheng}},
  \bibinfo {author} {\bibfnamefont {Bingqing}\ \bibnamefont {Zhu}}, \bibinfo
  {author} {\bibfnamefont {Xiankai}\ \bibnamefont {Sun}}, \ and\ \bibinfo
  {author} {\bibfnamefont {Hon~Ki}\ \bibnamefont {Tsang}},\ }\bibfield  {title}
  {\enquote {\bibinfo {title} {{Hyperuniform Disordered Network Polarizers}},}\
  }\href {\doibase 10.1109/JSTQE.2016.2528125} {\bibfield  {journal} {\bibinfo
  {journal} {IEEE J. Sel. Top. Quantum Electron.}\ }\textbf {\bibinfo {volume}
  {22}},\ \bibinfo {pages} {288--294} (\bibinfo {year} {2016})}\BibitemShut
  {NoStop}%
\bibitem [{\citenamefont {Degl'Innocenti}\ \emph {et~al.}(2016)\citenamefont
  {Degl'Innocenti}, \citenamefont {Shah}, \citenamefont {Masini}, \citenamefont
  {Ronzani}, \citenamefont {Pitanti}, \citenamefont {Ren}, \citenamefont
  {Jessop}, \citenamefont {Tredicucci}, \citenamefont {Beere},\ and\
  \citenamefont {Ritchie}}]{DeglInnocenti2016}%
  \BibitemOpen
  \bibfield  {author} {\bibinfo {author} {\bibfnamefont {R.}~\bibnamefont
  {Degl'Innocenti}}, \bibinfo {author} {\bibfnamefont {Y.~D.}\ \bibnamefont
  {Shah}}, \bibinfo {author} {\bibfnamefont {L.}~\bibnamefont {Masini}},
  \bibinfo {author} {\bibfnamefont {A.}~\bibnamefont {Ronzani}}, \bibinfo
  {author} {\bibfnamefont {A.}~\bibnamefont {Pitanti}}, \bibinfo {author}
  {\bibfnamefont {Y.}~\bibnamefont {Ren}}, \bibinfo {author} {\bibfnamefont
  {D.~S.}\ \bibnamefont {Jessop}}, \bibinfo {author} {\bibfnamefont
  {A.}~\bibnamefont {Tredicucci}}, \bibinfo {author} {\bibfnamefont {H.~E.}\
  \bibnamefont {Beere}}, \ and\ \bibinfo {author} {\bibfnamefont {D.~A.}\
  \bibnamefont {Ritchie}},\ }\bibfield  {title} {\enquote {\bibinfo {title}
  {{Hyperuniform disordered terahertz quantum cascade laser}},}\ }\href
  {\doibase 10.1038/srep19325} {\bibfield  {journal} {\bibinfo  {journal} {Sci.
  Rep.}\ }\textbf {\bibinfo {volume} {6}},\ \bibinfo {pages} {19325} (\bibinfo
  {year} {2016})}\BibitemShut {NoStop}%
\bibitem [{\citenamefont {{De Rosa}}\ \emph {et~al.}(2015)\citenamefont {{De
  Rosa}}, \citenamefont {Auriemma}, \citenamefont {Diletto}, \citenamefont {{Di
  Girolamo}}, \citenamefont {Malafronte}, \citenamefont {Morvillo},
  \citenamefont {Zito}, \citenamefont {Rusciano}, \citenamefont {Pesce},\ and\
  \citenamefont {Sasso}}]{DeRosa2015}%
  \BibitemOpen
  \bibfield  {author} {\bibinfo {author} {\bibfnamefont {C.}~\bibnamefont {{De
  Rosa}}}, \bibinfo {author} {\bibfnamefont {F.}~\bibnamefont {Auriemma}},
  \bibinfo {author} {\bibfnamefont {C.}~\bibnamefont {Diletto}}, \bibinfo
  {author} {\bibfnamefont {R.}~\bibnamefont {{Di Girolamo}}}, \bibinfo {author}
  {\bibfnamefont {a.}~\bibnamefont {Malafronte}}, \bibinfo {author}
  {\bibfnamefont {P.}~\bibnamefont {Morvillo}}, \bibinfo {author}
  {\bibfnamefont {G.}~\bibnamefont {Zito}}, \bibinfo {author} {\bibfnamefont
  {G.}~\bibnamefont {Rusciano}}, \bibinfo {author} {\bibfnamefont
  {G.}~\bibnamefont {Pesce}}, \ and\ \bibinfo {author} {\bibfnamefont
  {a.}~\bibnamefont {Sasso}},\ }\bibfield  {title} {\enquote {\bibinfo {title}
  {{Toward hyperuniform disordered plasmonic nanostructures for reproducible
  surface-enhanced Raman spectroscopy}},}\ }\href {\doibase 10.1039/C4CP06024E}
  {\bibfield  {journal} {\bibinfo  {journal} {Phys. Chem. Chem. Phys.}\
  }\textbf {\bibinfo {volume} {17}},\ \bibinfo {pages} {8061--8069} (\bibinfo
  {year} {2015})}\BibitemShut {NoStop}%
\bibitem [{\citenamefont {Leseur}\ \emph {et~al.}(2016)\citenamefont {Leseur},
  \citenamefont {Pierrat},\ and\ \citenamefont {Carminati}}]{Leseur2016}%
  \BibitemOpen
  \bibfield  {author} {\bibinfo {author} {\bibfnamefont {O.}~\bibnamefont
  {Leseur}}, \bibinfo {author} {\bibfnamefont {R.}~\bibnamefont {Pierrat}}, \
  and\ \bibinfo {author} {\bibfnamefont {R.}~\bibnamefont {Carminati}},\
  }\bibfield  {title} {\enquote {\bibinfo {title} {{High-density hyperuniform
  materials can be transparent}},}\ }\href {\doibase 10.1364/OPTICA.3.000763}
  {\bibfield  {journal} {\bibinfo  {journal} {Optica}\ }\textbf {\bibinfo
  {volume} {3}},\ \bibinfo {pages} {763} (\bibinfo {year} {2016})}\BibitemShut
  {NoStop}%
\bibitem [{\citenamefont {Florescu}\ \emph {et~al.}(2013)\citenamefont
  {Florescu}, \citenamefont {Steinhardt},\ and\ \citenamefont
  {Torquato}}]{PhysRevB.87.165116}%
  \BibitemOpen
  \bibfield  {author} {\bibinfo {author} {\bibfnamefont {Marian}\ \bibnamefont
  {Florescu}}, \bibinfo {author} {\bibfnamefont {Paul~J.}\ \bibnamefont
  {Steinhardt}}, \ and\ \bibinfo {author} {\bibfnamefont {Salvatore}\
  \bibnamefont {Torquato}},\ }\bibfield  {title} {\enquote {\bibinfo {title}
  {Optical cavities and waveguides in hyperuniform disordered photonic
  solids},}\ }\href {\doibase 10.1103/PhysRevB.87.165116} {\bibfield  {journal}
  {\bibinfo  {journal} {Phys. Rev. B}\ }\textbf {\bibinfo {volume} {87}},\
  \bibinfo {pages} {165116} (\bibinfo {year} {2013})}\BibitemShut {NoStop}%
\bibitem [{\citenamefont {Amoah}\ and\ \citenamefont
  {Florescu}(2015)}]{Amoah2015}%
  \BibitemOpen
  \bibfield  {author} {\bibinfo {author} {\bibfnamefont {Timothy}\ \bibnamefont
  {Amoah}}\ and\ \bibinfo {author} {\bibfnamefont {Marian}\ \bibnamefont
  {Florescu}},\ }\bibfield  {title} {\enquote {\bibinfo {title} {{High-Q
  optical cavities in hyperuniform disordered materials}},}\ }\href {\doibase
  10.1103/PhysRevB.91.020201} {\bibfield  {journal} {\bibinfo  {journal} {Phys.
  Rev. B}\ }\textbf {\bibinfo {volume} {91}},\ \bibinfo {pages} {020201}
  (\bibinfo {year} {2015})},\ \Eprint {http://arxiv.org/abs/1504.07055}
  {arXiv:1504.07055} \BibitemShut {NoStop}%
\bibitem [{\citenamefont {Tsitrin}\ \emph {et~al.}(2015)\citenamefont
  {Tsitrin}, \citenamefont {Williamson}, \citenamefont {Amoah}, \citenamefont
  {Nahal}, \citenamefont {Chan}, \citenamefont {Florescu},\ and\ \citenamefont
  {Man}}]{Band2016}%
  \BibitemOpen
  \bibfield  {author} {\bibinfo {author} {\bibfnamefont {Samuel}\ \bibnamefont
  {Tsitrin}}, \bibinfo {author} {\bibfnamefont {Eric~Paul}\ \bibnamefont
  {Williamson}}, \bibinfo {author} {\bibfnamefont {Timothy}\ \bibnamefont
  {Amoah}}, \bibinfo {author} {\bibfnamefont {Geev}\ \bibnamefont {Nahal}},
  \bibinfo {author} {\bibfnamefont {Ho~Leung}\ \bibnamefont {Chan}}, \bibinfo
  {author} {\bibfnamefont {Marian}\ \bibnamefont {Florescu}}, \ and\ \bibinfo
  {author} {\bibfnamefont {Weining}\ \bibnamefont {Man}},\ }\bibfield  {title}
  {\enquote {\bibinfo {title} {{Unfolding the band structure of non-crystalline
  photonic band gap materials.}}}\ }\href {\doibase 10.1038/srep13301}
  {\bibfield  {journal} {\bibinfo  {journal} {Sci. Rep.}\ }\textbf {\bibinfo
  {volume} {5}},\ \bibinfo {pages} {13301} (\bibinfo {year}
  {2015})}\BibitemShut {NoStop}%
\bibitem [{\citenamefont {Novotny}(2006)}]{Novotny2006}%
  \BibitemOpen
  \bibfield  {author} {\bibinfo {author} {\bibfnamefont {Lukas}\ \bibnamefont
  {Novotny}},\ }\href {\doibase 10.1016/S1748-0132(06)70120-7} {\emph {\bibinfo
  {title} {{Principles of Nano-Optics}}}}\ (\bibinfo  {publisher} {Cambridge
  University Press},\ \bibinfo {address} {Cambridge, UK},\ \bibinfo {year}
  {2006})\ \Eprint {http://arxiv.org/abs/arXiv:1011.1669v3}
  {arXiv:arXiv:1011.1669v3} \BibitemShut {NoStop}%
\end{thebibliography}

%

\end{document}